\documentclass[useAMS,natbib,multicolumn,subfigure]{mn2e}
\usepackage[latin1]{inputenc}  
\usepackage{graphicx,amsmath,xcolor}
\usepackage{amssymb}
\usepackage{indentfirst}
\usepackage{hyperref}
\usepackage{graphicx,pst-all}
\usepackage{tikz}
\usepackage{layout}
\usepackage{fancyhdr}
\def\msol{\hbox{\kern 0.20em $M_\odot$}}

\newcommand{\lsol}{\hbox{\kern 0.20em $L_\odot$}}
\newcommand{\g}{\hbox{\kern 0.20em g}}
\newcommand{\gmu}{\hbox{\kern 0.20em g$^{-1}$}}
\newcommand{\kg}{\hbox{\kern 0.20em kg}}
\newcommand{\pc}{\hbox{\kern 0.20em pc}}
\newcommand{\mum}{\hbox{\kern 0.20em $\mu$m}}
\newcommand{\mumd}{\hbox{\kern 0.20em $\mu$m$^{-2}$}}
\newcommand{\cm}{\hbox{\kern 0.20em cm}}
\newcommand{\m}{\hbox{\kern 0.20em m}}
\newcommand{\km}{\hbox{\kern 0.20em km}}
\newcommand{\nm}{\hbox{\kern 0.20em nm}}
\newcommand{\s}{\hbox{\kern 0.20em s}}
\newcommand{\h}{\hbox{\kern 0.20em h}}
\newcommand{\smu}{\hbox{\kern 0.20em s$^{-1}$}}
\newcommand{\srmu}{\hbox{\kern 0.20em sr$^{-1}$}}
\newcommand{\smd}{\hbox{\kern 0.20em s$^{-2}$}}
\newcommand{\an}{\hbox{\kern 0.20em an}}
\newcommand{\anmu}{\hbox{\kern 0.20em an$^{-1}$}}
\newcommand{\yr}{\hbox{\kern 0.20em yr}}
\newcommand{\yrmu}{\hbox{\kern 0.20em yr$^{-1}$}}
\newcommand{\Myr}{\hbox{\kern 0.20em Myr}}
\newcommand{\Mymu}{\hbox{\kern 0.20em Myr$^{-1}$}}
\newcommand{\K}{\hbox{\kern 0.20em K}}
\newcommand{\pcmu}{\hbox{\kern 0.20em pc$^{-1}$}}
\newcommand{\pcmd}{\hbox{\kern 0.20em pc$^{-2}$}}
\newcommand{\pcmt}{\hbox{\kern 0.20em pc$^{-3}$}}
\newcommand{\kms}{\hbox{\kern 0.20em km\kern 0.20em s$^{-1}$}}
\newcommand{\kmpd}{\hbox{\kern 0.20em km$^{2}$}}
\newcommand{\kpc}{\hbox{\kern 0.20em kpc}}
\newcommand{\cms}{\hbox{\kern 0.20em cm\kern 0.20em s$^{-1}$}}
\newcommand{\erg}{\hbox{\kern 0.20em erg}}
\newcommand{\ergs}{\hbox{\kern 0.20em erg}}
\newcommand{\cmpd}{\hbox{\kern 0.20em cm$^2$}}
\newcommand{\cmmd}{\hbox{\kern 0.20em cm$^{-2}$}}
\newcommand{\cmms}{\hbox{\kern 0.20em cm$^{-6}$}}
\newcommand{\cmpt}{\hbox{\kern 0.20em cm$^3$}}
\newcommand{\cmmt}{\hbox{\kern 0.20em cm$^{-3}$}}
\newcommand{\mpd}{\hbox{\kern 0.20em m$^2$}}
\newcommand{\mmd}{\hbox{\kern 0.20em m$^{-2}$}}
\newcommand{\mpt}{\hbox{\kern 0.20em m$^3$}}
\newcommand{\mmt}{\hbox{\kern 0.20em m$^{-3}$}}
\newcommand{\mujy}{\hbox{\kern 0.20em $\mu$Jy}}
\newcommand{\mjy}{\hbox{\kern 0.20em mJy}}
\newcommand{\Mj}{\hbox{\kern 0.20em MJy}}
\newcommand{\jy}{\hbox{\kern 0.20em Jy}}
\newcommand{\ghz}{\hbox{\kern 0.20em GHz}}
\newcommand{\G}{\hbox{\kern 0.20em G}}
\newcommand{\muG}{\hbox{\kern 0.20em $\mu$G}}

\newcommand{\thco}{\hbox{${}^{13}$CO}}

\newcommand{\ceio}{\hbox{C${}^{18}$O}}

\newcommand{\htwo}{\hbox{H${}_2$}}

\usetikzlibrary{decorations}
\pgfdeclaredecoration{ddbond}{initial}
{
\state{initial}[width=4pt]
{
\pgfpathlineto{\pgfpoint{4pt}{0pt}}
\pgfpathmoveto{\pgfpoint{2pt}{2pt}}
\pgfpathlineto{\pgfpoint{4pt}{2pt}}
\pgfpathmoveto{\pgfpoint{4pt}{0pt}}
}
\state{final}
{
\pgfpathlineto{\pgfpointdecoratedpathlast}
}
}
\tikzset{lddbond/.style={decorate,decoration=ddbond}}
\tikzset{rddbond/.style={decorate,decoration={ddbond,mirror}}}
\title[Peptide links in protostellar shocks]{Molecules with a peptide link in protostellar shocks: a comprehensive study of L1157}

\author[Edgar Mendoza et al.]
{Edgar~Mendoza $^{1,2,3}\thanks{E-mail:emendoza@astro.ufrj.br}$,
B. ~Lefloch$^{2,3}$,
A. ~L\'opez-Sepulcre$^{2,3}$,
C. ~Ceccarelli$^{2,3}$, C. Codella$^4$,
\newauthor 	%
H. M. ~Boechat-Roberty$^1$~\&~R. ~Bachiller$^5$
\\
\\
$^{1}$Observat\'orio do Valongo, Universidade Federal do Rio de Janeiro - UFRJ, Rio de
Janeiro, Brazil\\
$^{2}$Univ. Grenoble Alpes, IPAG, F-38000 Grenoble, France\\
$^3$CNRS, IPAG, F-38000 Grenoble, France \\
$^{4}$INAF, Osservatorio Astrofisico di Arcetri, Largo Enrico Fermi 5, I-50125 Firenze, Italy\\
$^{5}$IGN, Observatorio Astron\'omico Nacional, Calle Alfonso XIII, 3. 28004 Madrid, Spain\\
}
\begin{document}
\date{Received / Accepted}
\pagerange{\pageref{firstpage}--\pageref{lastpage}} \pubyear{2014}
\maketitle
\label{firstpage}
\begin{abstract}
Interstellar molecules with a peptide link -NH-C(=O)-, like formamide (NH$_2$CHO), acetamide (NH$_2$COCH$_3$) and isocyanic acid (HNCO) are particularly interesting for their potential role in pre-biotic chemistry. We have studied their emission in the protostellar shock regions L1157-B1 and L1157-B2, with the IRAM 30m telescope, as part of the ASAI Large Program.  Analysis of the line profiles shows that the emission arises from the outflow cavities associated with B1 and B2. Molecular abundance of~$\approx~(0.4-1.1)\times 10^{-8}$  and $(3.3-8.8)\times 10^{-8}$ are derived for formamide and isocyanic acid, respectively, from a simple  rotational diagram analysis. Conversely, NH$_2$COCH$_3$ was not detected down to a relative abundance of a few $\leq 10^{-10}$. B1 and B2 appear to be among the richest Galactic sources of HNCO and NH$_2$CHO molecules.
 A tight linear correlation between their abundances is observed, suggesting that the two species are chemically related. Comparison with astrochemical models favours molecule formation on ice grain mantles, with NH$_2$CHO generated  from hydrogenation of HNCO.
\end{abstract}

\section{Introduction}
The search for pre-biotic molecules represents a challenge for knowledge, which lead us to investigate reactions in physical and chemistry conditions radically different from those observed on Earth. Molecular species with a peptide link (-NH-C(=O)-) play an important role in the synthesis of amino acids and proteins, which are essential to living systems. 
A few species with the peptide link have  been detected in the interstellar medium, starting with HNCO itself, which was detected in many  prestellar and protostellar regions (Bisschop et al. 2007; Marcelino et~al. 2009; Rodr\'iguez-Fern\'andez et~al. 2010).  The second molecule is formamide, which was detected for the first time in space towards Sgr B2 by Rubin et~al. (1971). Since then, it has been observed in several Galactic environments such as molecular clouds, high-mass star forming regions and comets (Bockel\'ee-Morvan et~al. 2000; Nummelin et~al. 2000; Bisschop et~al. 2007; Motiyenko et~al. 2012).  It has also been tentatively assigned within the solid phase on icy grains toward the protostellar objects NGC 7538 IRS9 and W33A, based on ISO-SWS spectra (Schutte et~al. 1999; Raunier et~al. 2004). Recently, Kahane et~al. (2013) reported its detection in the envelope of the solar-type protostar IRAS16293-2422, proving that it can be easily synthesized also in solar-type environments. One of the studies revealing the pre-biotic potential of formamide, show that formamide solutions heated and irradiated lead to the synthesis of nucleobases, compounds present in DNA and RNA (Saladino et~al. 2007, 2012;  Barks et~al. 2010). Such a result, in the primordial soup scenario, constitutes an important step in the understanding of the origin of life. Acetamide (NH$_2$COCH$_3$) is the next most simple amide after formamide, and so far the largest interstellar molecule with a peptide bond (Hollis et~al. 2006).  Halfen, Ilyushin \& Ziurys~(2011) have shown that acetamide is one of the most abundant complex organic species in Sgr B2(N), and could be a possible source of larger peptide molecules.

The similarities noticed by  Bockel\'ee-Morvan et~al. (2000) in the  molecular content of Comet Hale-Bopp  and protostellar shock regions, suggest the possibility that formamide and, more generally, molecules with a peptide bond, could be quite abundant in shock regions. This  led us to search for isocyanic acid, formamide and acetamide,  in the chemically active outflow powered by L1157-mm,
a low-mass Class 0 protostar  ($\sim 4 L_{\odot}$), located at $\sim 250$ pc  (Looney, Tobin \& Kwon  2007). The outflow interacts strongly with the ambient gas through shocks, as revealed in radio line emission (Bachiller, Mart\'in-Pintado \& Fuente 1993) and in the pure rotational lines of \htwo\ in the mid-infrared (Nisini et~al. 2010). The ejected material interacts with the ambient gas through shock fronts, which heat the gas and trigger endothermic reactions, sputtering and desorption processes from dust grains. Towards the southern lobe, two bowshocks are detected, identified as L1157-B1 and L1157-B2 (hereafter B1 and B2), associated with subsequent events in the mass-loss phenomenon, with dynamical ages of $\sim$ 2000 and 4000 yr for B1 and B2, respectively (Gueth, Guilloteau \& Bachiller 1996). Protostellar bowshock B1 displays the brightest molecular line emission in the Southern outflow lobe (Bachiller et~al. 2001). As part of the IRAM 30m Large Program ASAI dedicated to astrochemical surveys of solar-type star forming regions, we have carried out an unbiased spectral survey of the molecular line emission towards  B1 between 80 and 350 GHz (Lefloch et~al. in preparation). The unprecedented sensitivity of the survey reveals the copious complex chemistry at work in the bowshock (Codella et~al. 2012; Podio et~al. 2014). We have obtained complementary observations towards the older protostellar bowshock B2.

The paper is organized as follows.
After presenting the observations and the data reduction (Section 2), we discuss the physical properties of formamide, isocyanic acid and acetamide in Sections 3, 4, and 5, respectively. In Section 6, we compare their properties with possible formation pathways. We summarize our results in Section 7.

\section{Observations and data reduction}
The observations of B1 were carried out with the IRAM-30m telescope at Pico Veleta (Spain), as part of the ASAI Large Program. The observed position in B1 is $\alpha_{J 2000} =$ 20$^{\text h}$ 39$^{\text m}$ 10.$^{\text s}$2, $\delta_{J 2000} =$ +68$^{\circ}$ 01$^{\prime}$ 10$^{\prime\prime}$. The survey was acquired during several runs in 2011 and 2012, using  the broad-band EMIR receivers at 3~mm (80 -- 116 GHz), 2~mm (128 -- 173 GHz), 1.3~mm (200 -- 320 GHz), and 0.8~mm (329 -- 350 GHz). Fast Fourier Transform Spectrometers and the WILMA autocorrelator were connected to the EMIR receivers, providing a spectral resolution of 195 kHz and 2 MHz, respectively. The final kinematic resolution of the data was degraded to $1\kms$. The observations were carried out in Wobbler Switching Mode, with a throw of $3^{\prime}$, in order to ensure a flat baseline across the spectral bandwith observed (4 GHz to 8 GHz, depending on the receiver).
Complementary observations of B2 of the band 83.5 -- 107 GHz were carried out in December 2012 and December 2013 with the IRAM 30m telescope, towards  the nominal position of B2: $\alpha_{J 2000} =$ 20$^{\text h}$ 39$^{\text m}$ 12.$^{\text s}$6, $\delta_{J 2000} =$ +68$^{\circ}$ 00$^{\prime}$ 40$^{\prime\prime}$. We adopted the same instrumental setup as for the B1 observations.

The data reduction was performed using the GILDAS/CLASS90 package\footnote{http://www.iram.fr/IRAMFR/GILDAS/}. The line intensities are expressed in units of antenna temperature corrected for atmospheric attenuation and rearward losses ($T_a^{\star}$). For subsequent analysis, fluxes were expressed in main beam temperature units ($T_{mb}$).  We have adopted a calibration uncertainty of 10, 15 and 20~\% for the surveys at 3 and 2~mm, 1.3~mm and 0.8~mm, respectively. The telescope and receiver parameters (main-beam efficiency, Beff; forward efficiency, Feff; Half Power beam Width, HPBW) were taken from the IRAM webpage\footnote{http://www.iram.es/IRAMES/mainWiki/Iram30mEfficiencies}.

We have summarized the spectroscopic properties and the observational parameters of the molecular transitions detected in Table~\ref{tab:table1} and~\ref{tab:table2}.

\begin{table*}
\caption{Spectroscopic and observational parameters of the molecular transitions of NH$_2$CHO, HNCO and HCNO lines detected in L1157-B1. Integrated area, FWHM and velocity peak are obtained from a Gaussian fit to the line profile. Statistical uncertainties are given in brackets.
\label{tab:table1}}
\begin{tabular}{llcccrrc}
\hline
Transition  & Frequency & $A_{ij}$ & $E_u$   & HPBW & $\int T_{mb}dv$  & FWHM & $V$ \\
$J_{k_a^{\prime}k_c^{\prime}}^{\prime} \rightarrow J_{k_ak_c}$ & MHz & $10^{-5}\smu$ & K  & $\arcsec$ &   mK $\kms$  & $\kms$ & $\kms$ \\
\hline
NH$_2$CHO &  &  &  &  &  &   & \\
$4_{1,4} \rightarrow 3_{1,3}$          & 81693.446 & 3.46 & 12.8& 30.1 &  93(8)   & 5.4(7) & 0.6(2)  \\
$4_{0,4} \rightarrow  3_{0,3}$          & 84542.330 & 4.09 & 10.2& 29.1 & 118(5)   & 6.1(3) & 0.9(2)  \\
$4_{2,3} \rightarrow  3_{2,2}$          & 84807.795 & 3.09 & 22.1& 29.0 &  19(3)   & 3.4(8) & 0.8(3)  \\
$4_{2,2} \rightarrow  3_{2,1}$          & 85093.272 & 3.13 & 22.1& 28.9 &  22(4)   & 5.0(1)   & 0.5(6)  \\
$4_{1,3} \rightarrow 3_{1,2}$           & 87848.873 & 4.30 & 13.5& 28.0 & 112(4)   & 5.5(4) & 0.2(2)  \\
$3_{2,2} \rightarrow 4_{1,3}$ $^{\dag}$ & 93871.691 & 0.04 & 18.0& 26.2 & 203.9(8) & 5.0(2) & 4.7(1)  \\
$5_{1,5} \rightarrow 4_{1,4}$          & 102064.267& 7.06 & 17.7& 24.1 & 107(6)   & 6.5(5) & 0.6(3)  \\
$5_{0,5} \rightarrow 4_{0,4}$          & 105464.219& 8.11 & 15.2& 23.3 & 87(7)    & 5.1(6) & 0.4(3)  \\
$5_{2,4} \rightarrow 4_{2,3}$          & 105972.599& 6.92 & 27.2& 23.2 & 35(6)    & 3.0(1)   & 0.1(3)  \\
$5_{2,3}\rightarrow 4_{2,2}$           & 106541.679& 7.03 & 27.2& 23.1 & 41(1)    & 6.0(2)   & 0.4(6)  \\
$5_{1,4}\rightarrow 4_{1,3}$           & 109753.503& 8.78 & 18.8& 22.4 & 85(7)    & 5.2(5) & 0.4(2)  \\
$6_{1,5}\rightarrow  5_{1,4}$          & 131617.902& 15.6 & 25.1& 18.7 &  61(10)  & 2.9(6) & 1.1(3)   \\
$7_{1,7}\rightarrow  6_{1,6}$          & 142701.325& 20.2 & 30.4& 17.2 &  35(4)   & 1.9(5) & 0.8(2)   \\
$7_{0,7}\rightarrow  6_{0,6}$          & 146871.475& 22.5 & 28.3& 16.7 &  49(10)  & 3.7(7) & 0.7(4)   \\
$7_{1,6}\rightarrow  6_{1,5}$          & 153432.176& 25.1 & 32.5& 16.0 &  45(15)  & 2.5(6) & 1.4(3)   \\
$5_{1,5}\rightarrow  4_{0,4}$          & 156835.561& 0.92 & 17.7& 15.7 & 31(10)   & 1.8(7) & 2.3(3)   \\
$10_{1,10}\rightarrow 9_{1,9}$         & 203335.761& 60.3 & 56.8& 12.1 & 23(10)   & 3.0(2)   & 1.5(7)     \\
$10_{0,10}\rightarrow 9_{0,9}$         & 207679.189& 64.7 & 55.3& 11.8 & 35(10)   & 3.0(1)   & 1.6(4)    \\
$10_{2,9}\rightarrow 9_{2,8}$          & 211328.960& 65.6 & 67.8& 11.6 & 46(10)   & 4.0(1)   & 2.1(8)    \\
$10_{3,8}\rightarrow  9_{3,7}$         & 212572.837& 63.3 & 82.9& 11.6 & 23(5)    & 1.0(1)   & 2.0(4)   \\
$10_{1,9}\rightarrow 9_{1,8}$          & 218459.213& 74.8 & 60.8& 11.3 & 52(10)   & 5.0(1)   & 0.7(9)    \\
$10_{3,7}\rightarrow 11_{1,10}$        & 220538.374& 0.02 & 82.9& 11.2 & 32(5)    & 4.0(2)   & 2.1(7)    \\
$11_{0,11}\rightarrow  10_{0,10}$      & 227606.176& 2.10 & 66.3& 10.8 & 22(10)   & 2.0(1)   & 0.6(5)     \\
$11_{1,10}\rightarrow  10_{1,9}$       & 239952.354& 99.6 & 72.3& 10.2 & 59(10)   & 6.0(1)   & 1.4(5)    \\
$7_{2,6}\rightarrow 7_{0,7}$ $^{\ddag}$& 251567.075& 0.24 & 40.4& 9.8  & 76(10)   & 5.0(1)   & 1.5(5)    \\
\hline
HNCO &  &  &  &  &  &  &  \\
$4_{0,4}\rightarrow 3_{0,3}$    & 87925.237 & 0.88 & 10.55 & 27.9 & 828(7)  & 5.3(1) & 0.42(2) \\
$5_{0,5}\rightarrow  4_{0,4}$   & 109905.749 & 1.75 & 15.82 & 22.3 & 994(3) & 5.0(1) & 0.47(2) \\
$6_{0,6}\rightarrow  5_{0,5}$   & 131885.734 & 3.08 & 22.15 & 18.6 & 1019(20)& 4.5(1) & 0.58(3) \\
$7_{0,7}\rightarrow  6_{0,6}$   & 153865.086 & 4.94 & 29.53 & 15.9 & 1173(20)& 5.3(1) & 0.49(6) \\
$10_{0,10}\rightarrow 9_{0,9}$  & 219798.274 & 14.7 & 58.02 & 11.2 & 716(8) & 5.5(9) & 0.89(4)   \\
$12_{0,12}\rightarrow 11_{0,11}$ & 263748.625 & 25.6 & 82.28& 9.32 & 297(30)& 4.8(6) & 0.62(2)   \\
$13_{0,13} \rightarrow 12_{0,12}$& 285721.951 & 32.6 & 95.99& 8.61 & 281(20)& 7.0(1)   & 0.45(5)\\
$14_{0,14} \rightarrow 13_{0,13}$& 307693.905 & 40.9 &110.76& 7.99 & 257(20)& 4.5(7) & 0.52(0.47) \\
$15_{0,15} \rightarrow 14_{0,14}$& 329664.367 & 50.4 &126.58& 7.46 & 219(30)& 2.2(7) & 0.61(0.16)  \\
\hline
HCNO &  &  &  &  &  &  &  \\
$4 \rightarrow 3$    & 91751.320 & 3.8 & 11.0 & 26.8 & 44(5) & 3.9(5) & 1.5(2) \\
$5\rightarrow  4$   & 114688.382 & 7.67 & 16.5 & 21.4 & 22(8) & 2.1(6) & 1.1(4) \\
\hline
\end{tabular}
\begin{flushleft}
Number in parentheses represent uncertainties on the last digit.\\
$^{\dag}$  blended with CCS (7$_8$   $\rightarrow$  6$_7$) transition at 93870.107 MHz.\\
$^{\ddag}$ possibly blended with C$_2$H$_5$OH
\end{flushleft}
\end{table*}

\begin{table*}
\caption{Spectroscopic and observational parameters of the molecular transitions of   NH$_2$CHO and HNCO detected in L1157-B2.  Integrated area, FWHM and velocity peak are obtained from a Gaussian fit to the line profile. Statistical uncertainties are given in brackets.
\label{tab:table2}}
\begin{tabular}{llcccrrc}
\hline
Transition  & Frequency & $A_{ij}$ & $E_u$   & HPBW & $\int T_{mb}dv$  & FWHM & $V$ \\
$J_{k_a^{\prime}k_c^{\prime}}^{\prime} \rightarrow J_{k_ak_c}$ & MHz & $10^{-5}\smu$ & K  & $\arcsec$ &   mK $\kms$  & $\kms$ & $\kms$ \\
\hline
NH$_2$CHO &  &  &  &  &  &   & \\
$4_{0,4}\rightarrow  3_{0,3}$          & 84542.330 & 4.09 & 10.2& 29.1 & 185(20)  & 4.5(6)   & 1.4(3)\\
$4_{2,2}\rightarrow  3_{2,1}$          & 85093.272 & 3.13 & 22.1& 28.9 & 71(20)   & 4.0(1)     & 1.0(4) \\
$4_{1,3}\rightarrow 3_{1,2}$           & 87848.873 & 4.30 & 13.5& 28.0 & 185(10)  & 3.4(3)   & 1.6(1) \\
$3_{2,2}\rightarrow 4_{1,3}$ $^{\dag}$ & 93871.691 & 0.04 & 18.0& 26.2 & 122(5)   & 3.7(4)   & 6.9(1) \\
$5_{1,5} \rightarrow 4_{1,4}$          & 102064.267& 7.06 & 17.7& 24.1 & 119(10)  & 3.2(6)   & 1.9(2)\\
$5_{0,5} \rightarrow 4_{0,4}$          & 105464.219& 8.11 & 15.2& 23.3 & 168(10)  & 2.8(3)   & 1.6(1)\\
$5_{2,4} \rightarrow 4_{2,3}$          & 105972.599& 6.92 & 27.2& 23.2 & 59(20)   & 3.3(1)   & 0.7(4) \\
\hline
HNCO &  &  &  &  &  &  &  \\
$4_{0,4}\rightarrow 3_{0,3}$    & 87925.237  & 0.88 & 10.55 & 27.9 & 1263(14) & 3.3(1)  & 1.76(2)\\
$6_{0,6}\rightarrow  5_{0,5}$   & 131885.734 & 3.08 & 22.15 & 18.6 & 1650(28) &  2.7(1) & 1.39(7)\\
$10_{0,10}\rightarrow 9_{0,9}$  & 219798.274 & 14.7 & 58.02 & 11.2 & 650(22) &   2.5(1) & 1.88(4) \\
$12_{0,12}\rightarrow 11_{0,11}$ & 263748.625 & 25.6 & 82.28& 9.32 & 314(30)& 2.3(3) & 1.70(1) \\
\hline
\end{tabular}
\begin{flushleft}
Number in parentheses represent uncertainties on the last digit.\\
$^{\dag}$  blended with CCS (7$_8$   $\rightarrow$  6$_7$) transition at 93870.107 MHz.\\
\end{flushleft}
\end{table*}

\section{Formamide}

\subsection{Line emission}

Formamide displays a rich rotational line spectrum at millimeter wavelengths. Overall, we have detected twenty-five transitions towards B1  well above $3\sigma$. Due to the narrow spectral coverage of the B2 observations, seven transitions were detected towards the older shock. The line identification was performed using the CDMS\footnote{http://www.astro.uni-koeln.de/cgi-bin/cdmssearch} and JPL\footnote{http://spec.jpl.nasa.gov/ftp/pub/catalog/catform.html} molecular spectroscopy databases (M\"uller et~al. 2001; Pearson, Drouin \& Pickett 2005).  Our results are in agreement with Yamaguchi et~al. (2012), who reported the detection of four transitions of low energy between 80 and 115 GHz.

\begin{figure}
\begin{center}
\includegraphics[width=\columnwidth]{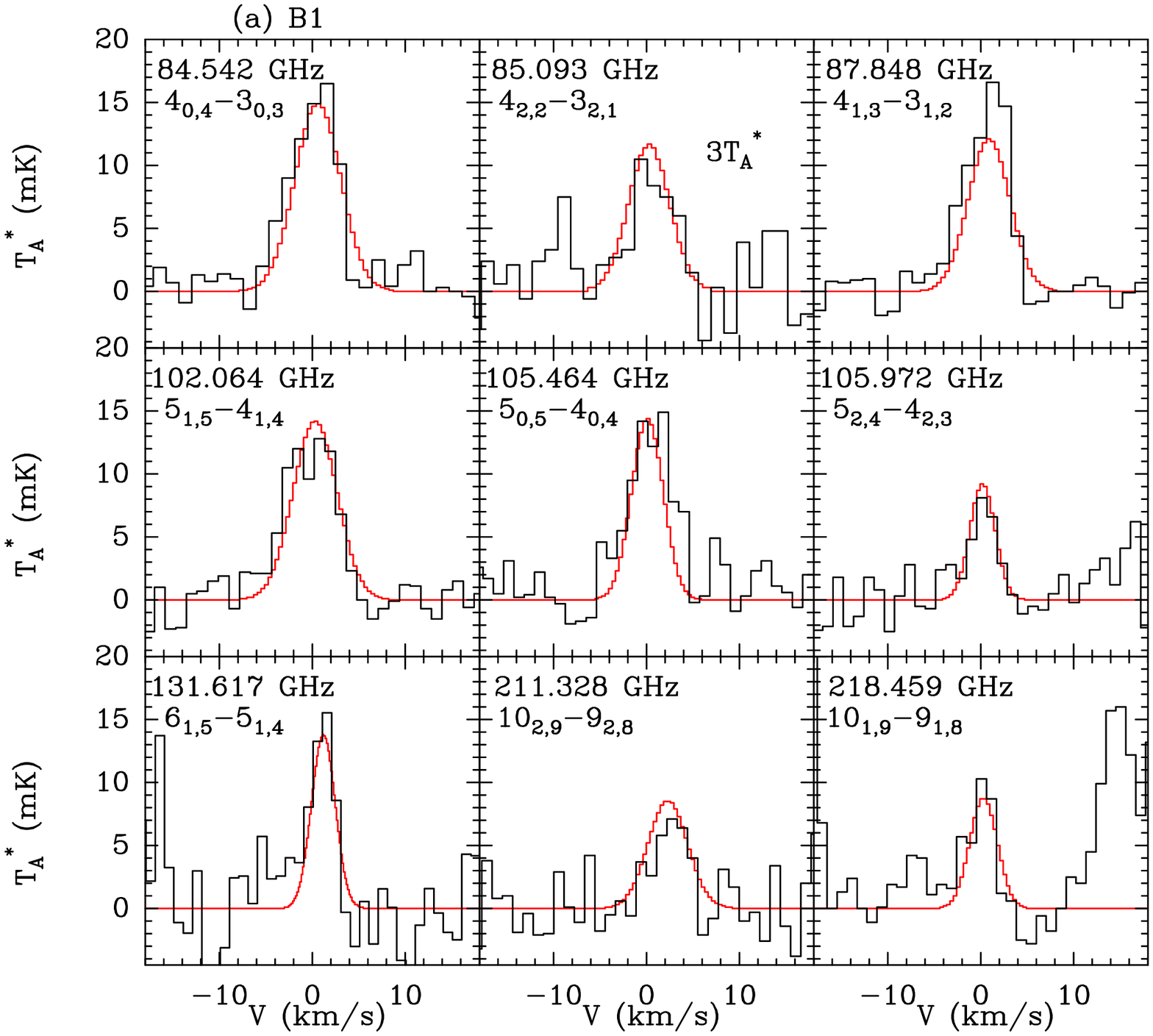}
\includegraphics[width=\columnwidth]{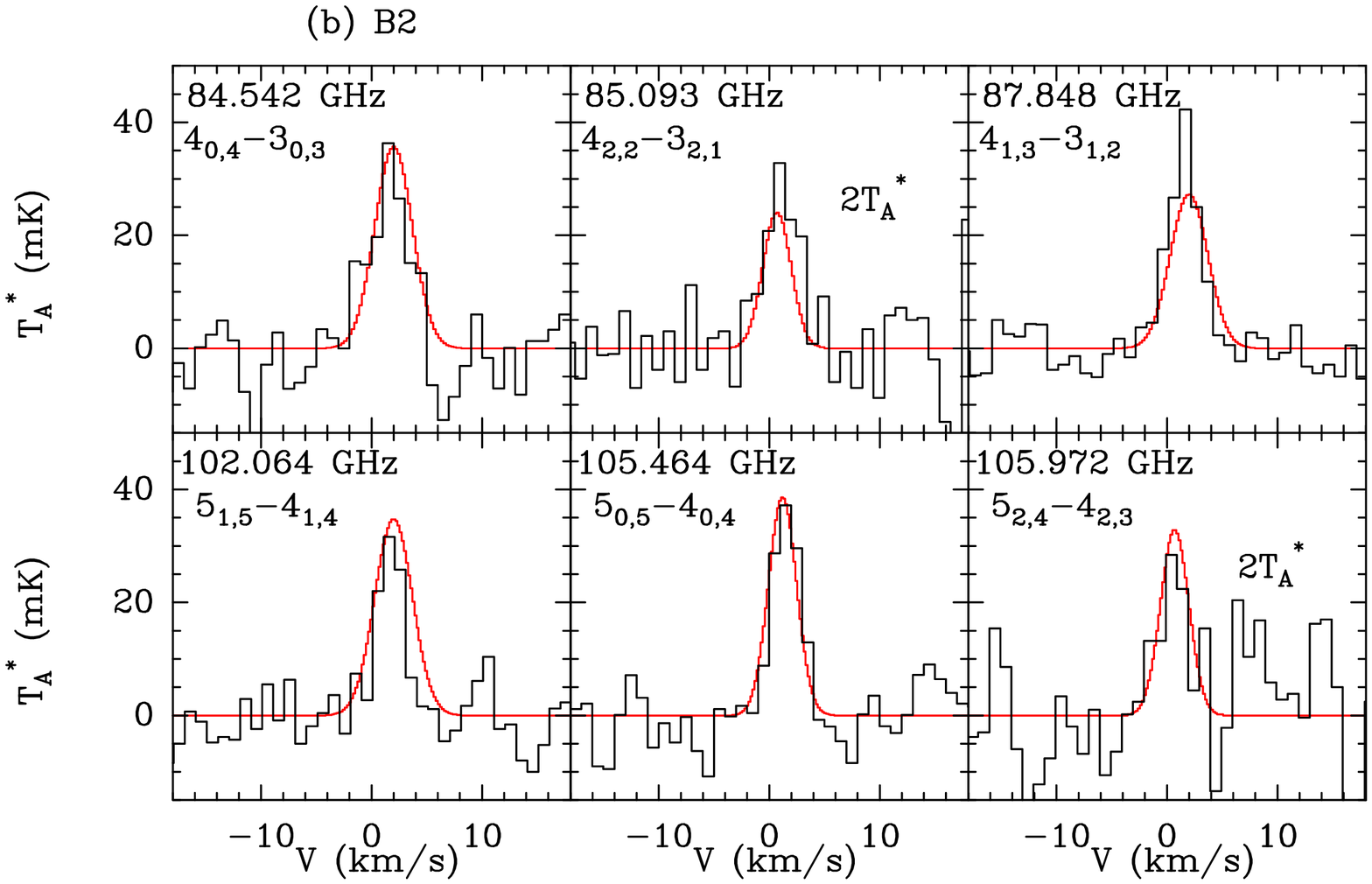}
{\caption{\small NH$_2$CHO lines observed towards the nominal positions of (a) B1 and (b) B2 (black histogram). The fit to the emission obtained from an LTE analysis is displayed in red.
\label{fig:figure1}
}}
\par\end{center}
\end{figure}

Two lines appear to be blended with transitions from other molecular species:
the $3_{2,2}-4_{1,3}$ line at 93871.699 MHz and the $7_{2,6}-7_{0,7}$ at 251567.075 MHz are blended with  CCS $7_8-6_7$ at 93870.091 MHz, and C$_2$H$_5$OH at 251566.455 MHz, respectively.  Both transitions are left aside in the rest of this work. Detected lines have an energy above the ground state ($E_u$) in the range $10-80\K$. At 3~mm (2~mm),  all the transitions with $E_u \leq 40\K$ and Einstein coefficient $(A_{ij}) \geq 2 \times 10^{-5}$ ($2\times 10^{-4}$)   are detected. This makes us confident about the unambiguous detection of formamide in B1 and B2. The observational parameters of the lines were derived from a Gaussian fit to the line profiles; they are summarized in Table~\ref{tab:table1} and~\ref{tab:table2}.
A montage of a few transitions detected towards B1 and B2 is displayed in Fig.~\ref{fig:figure1}.

Formamide emission is blueshifted with respect to the ambient cloud emission ($V = +2.6\ \kms$), and the emission peaks at about $V = +1\ \kms$. Linewidths are typically $ 5\kms$ for the low-excitation transitions ($E_u~<~40 \K$). We observe an excellent agreement between the line profiles of NH$_2$CHO and those of transitions from organic molecules such as CH$_3$OH, H$_2$CO and NH$_3$, which are thought to form on the ice mantles of dust grains (see e.g. Codella et~al. 2010; Fontani et~al. 2014).

\begin{figure}
\begin{center}
\includegraphics[width=0.97\columnwidth]{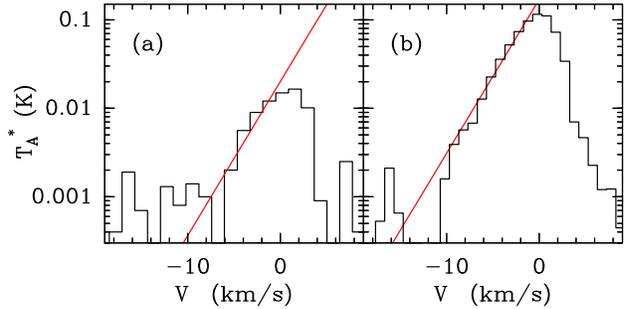}
{\caption{\small Montage of the (a) NH$_2$CHO 4$_{0,4}$--3$_{0,3}$ and (b) HNCO 4$_{0,4}$--3$_{0,3}$ transitions line profiles (black histogram). In both line profiles, the wing emission is fitted by an exponential function $\propto \exp(-|v/2.5|)$ (red straight line).
\label{fig:figure2}
}}
\par\end{center}
\end{figure}

We have analyzed the intensity-velocity distribution of formamide in B1, following the approach described in Lefloch et~al. (2012). The intensity-distribution as a function of velocity  $I(v)$ is displayed on a linear-logarithmic scale (see Fig.~\ref{fig:figure2}).
Lefloch et al. (2012)  have shown that the CO intensity-velocity distribution is the sum of three contributions from physically distinct components with specific excitation conditions: the jet-impact shock region against the B1 cavity (g1), the walls of the B1 cavity (g2), and a more extended component (g3) which the authors associated with the ejection responsible for the B2 cavity (g3). Each of these components displays an intensity-velocity distribution $I(v) \propto \exp(-|v/v_0|)$, where $v_0= 12.5, 4.4, 2.5\kms$ for g1, g2, g3, respectively. The contribution of each component to the emerging intensity appears as a straight line in Fig.~\ref{fig:figure2}. Each component is characterized by a specific slope over the full velocity range of emission, which allows to disentangle their respective contributions thanks to the difference of excitation conditions (see also Fig.~\ref{fig:figure1} in Lefloch et al. 2012).

Due to the low brightness of the formamide lines, this method could be applied only to the low-excitation transitions $4_{0,4}$--$3_{03}$ and $4_{1,3}$--$3_{1,2}$.
As shown in Fig.~\ref{fig:figure2}, the intensity-velocity distribution
$I(v)$ of the low-excitation transition is well fit by an exponential law $I(v)\propto \exp(-|v/v_0|)$, with $v_0=2.5\kms$. The emission of the low-excitation transitions of formamide appears to follow the same distribution as CO in the B2 cavity (g3), which suggests a similar origin. Several subsequent studies have confirmed this interpretation (Busquet et~al. 2014; G\'omez-Ruiz et~al. submitted).  We note that the profiles of the higher-excitation transitions display narrower linewidths $\approx 2\kms$, which suggests that the highest excitation gas is concentrated in a more restricted region.

\subsection{Physical conditions}
\begin{figure}
\begin{center}
\includegraphics[width=\columnwidth]{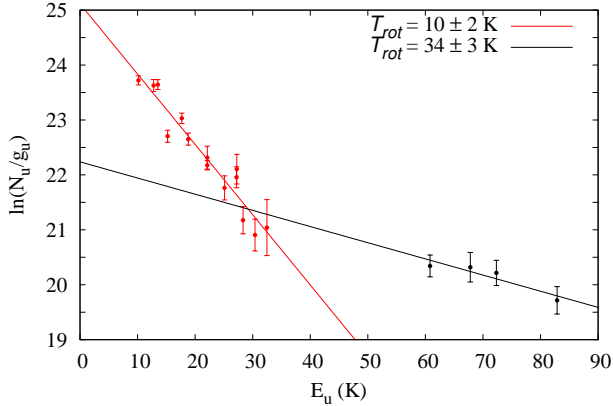}
{\caption{\small Rotational diagram of the NH$_2$CHO emission detected towards B1. The red and black straight lines display the fits to the low- ($T_{rot}= 10\K$) and high-excitation ($T_{rot}= 34\K$) components, respectively. 
\label{fig:figure3}
}}
\par\end{center}
\end{figure}

We have derived the excitation conditions of NH$_2$CHO under the hypothesis of local thermodynamic equilibrium (LTE).  Assuming the lines to be optically thin, the column density of molecules at a level $u$  and the rotational temperature ($T_{rot}$) are related
\begin{align}
\ln \frac{N_u}{g_u} = [\ln N_F - \ln Q(T_{rot})] - \frac{1}{T_{rot}}\left(\frac{E_u}{K}\right),
\end{align}
where $g_u$, $N_F$ and $Q(T_{rot})$ are the  degeneracy of the upper level ($u$), the total formamide column density and the formamide partition function, respectively. Using the above relation and adopting a source size of $18\arcsec$  for the formamide emitting region, similar to the size estimated for the CS and CO emissions (Lefloch et~al. 2012; G\'omez-Ruiz et~al. submitted), we have built the rotational diagram of formamide for B1 and B2, taking into account the total flux uncertainties. They are displayed in Figs.~\ref{fig:figure3} and \ref{fig:figure5}.

In the case of B1, a simple rotational diagram shows evidence for the presence of two components of different excitation (see Fig.~\ref{fig:figure3}): a component of low-excitation, with $T_{rot}\simeq~10\pm2\K$, and column density $N \rm (NH_2CHO)= 3.5\pm0.6\times 10^{12}\cmmd$, and a component of high-excitation, with $T_{rot}\simeq~34\pm3\K$, and column density $N\rm (NH_2CHO)= 1.7\pm0.4\times 10^{12}\cmmd$. Using CASSIS\footnote{http://cassis.irap.omp.eu} and adopting these physical conditions for formamide, we have computed for all the transitions the line profile expected (predicted) for each component.  As can be seen in Fig.~\ref{fig:figure1}, a very good agreement is obtained between the fits (in red) and the individual spectra (in black).

Based on the spectral signature of the low- and high- excitation transitions of formamide, we conclude that the two components are real and not a consequence of non-LTE level population (e.g. Goldsmith \& Langer 1999).  The low-excitation emission arises from the g3 component, a layer of cold gas with kinetic temperature $T_{kin}\simeq 23\K$ and column density $N\rm (\htwo)\approx 1.0\times 10^{21} \cmmd$.  From the analysis of the CO and HNCO emissions (see below), we propose that the high-excitation NH$_2$CHO emission arises from the walls of the B1 cavity, where the kinetic temperature is estimated $\approx 60$ -- $80\K$ (Lefloch et~al. 2012). Hence, we derive an abundance $\rm [NH_2CHO]= 3.5\pm0.6\times 10^{-9}$ and $1.7\pm0.4 \times 10^{-9}$  for the low- and high-excitation components, respectively.

Towards B2, we detected a single gas component with $T_{rot}= 10\pm3\K$ and column density $N\rm (NH_2CHO)= 5.9\pm1.4 \times 10^{12}\cmmd$. However, due to the narrow range of frequency and $E_u$ covered towards B2, we cannot exclude the presence of components of higher excitation. From the CO column density estimated by Bachiller \& P\'erez Guti\'errez (1997), we derive $\rm [NH_2CHO]\approx 1.1\pm0.2\times 10^{-8}$.

\section{Isocyanic acid}

\subsection{Line emission}
\begin{figure}
\begin{center}
\includegraphics[width=8cm,keepaspectratio]{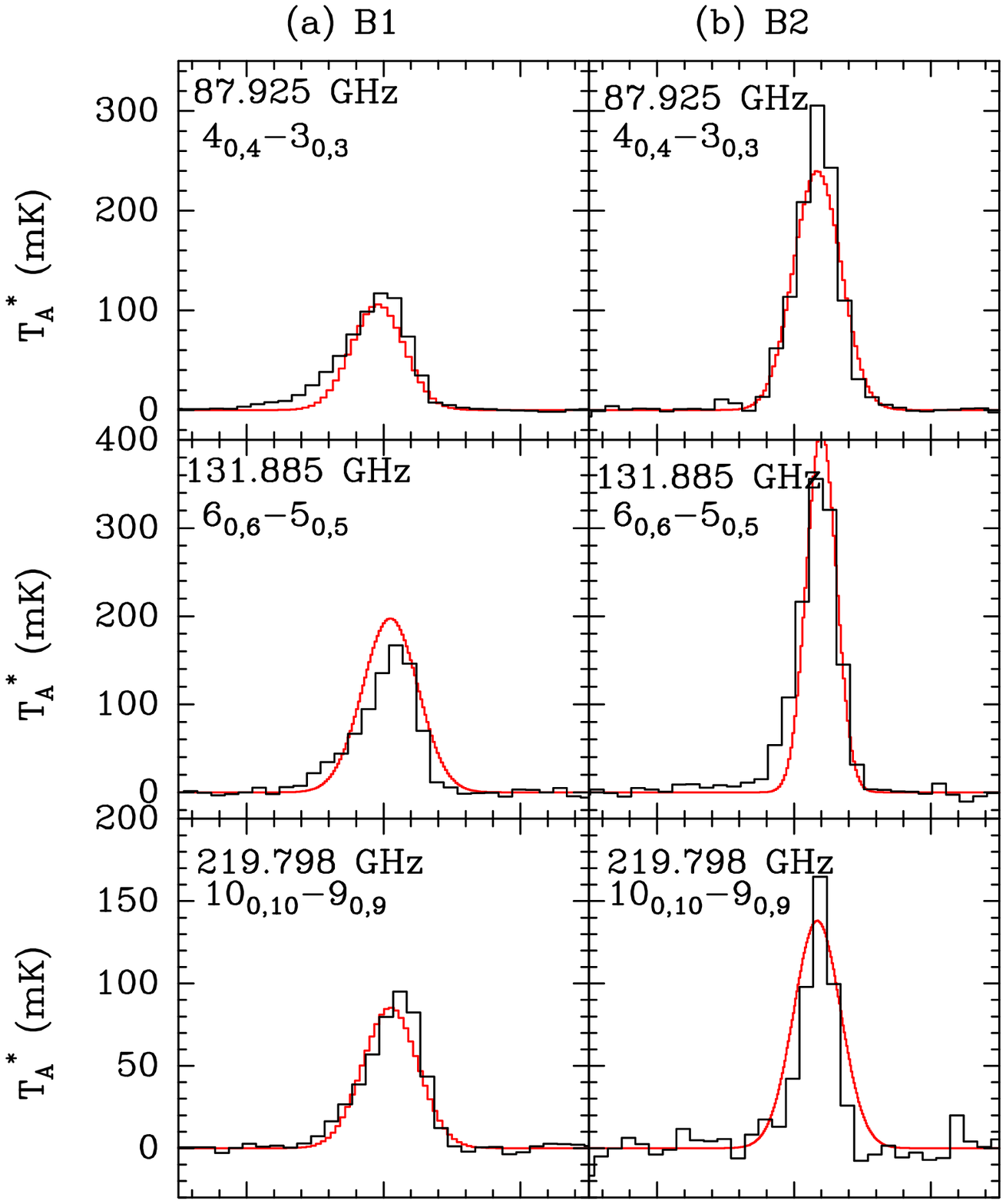}
\includegraphics[width=8cm,keepaspectratio]{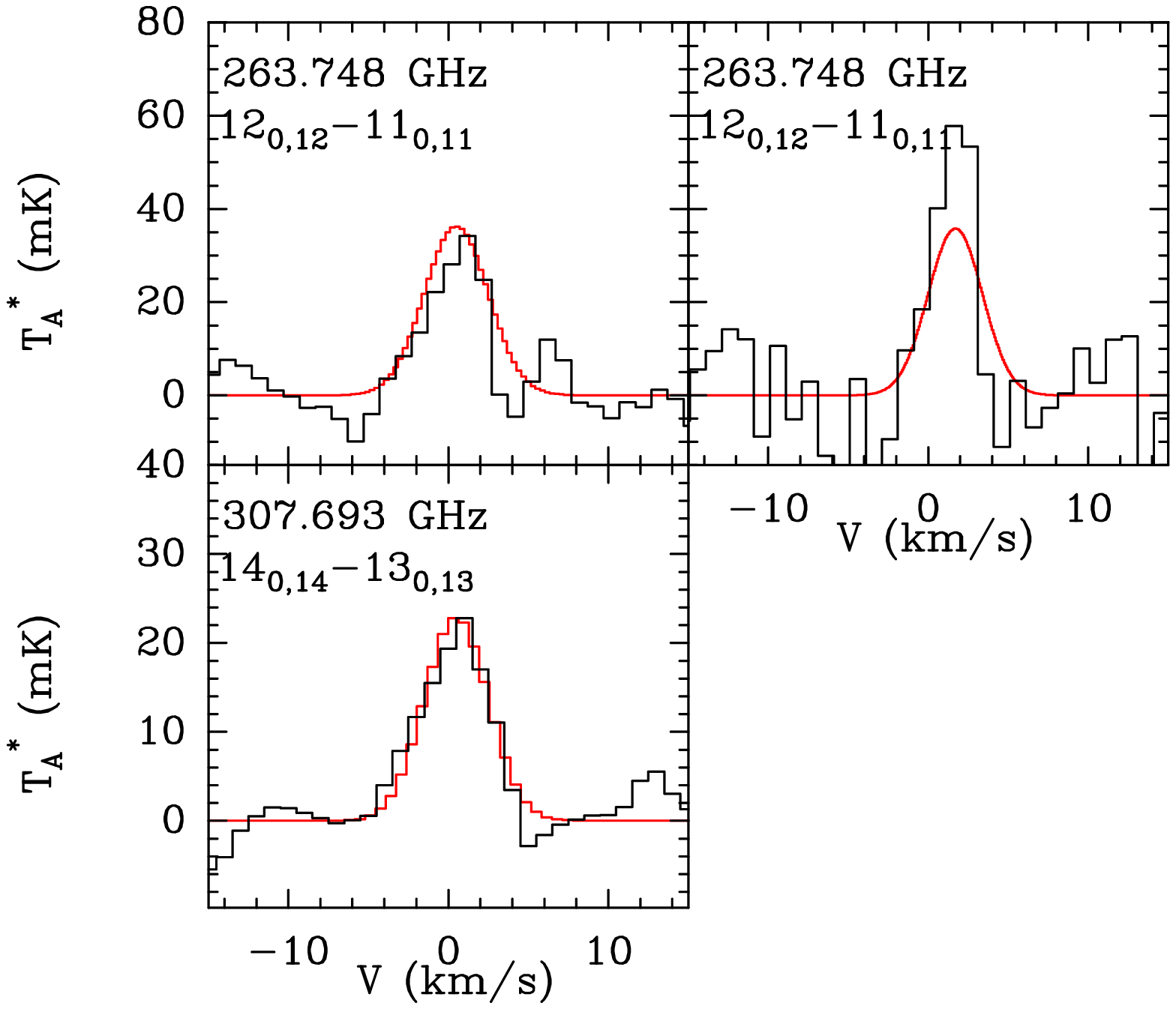}
{\caption{Montage of HNCO lines detected towards  (a)~B1 (left column) and (b)~B2 (right column). The LTE fit to the emission obtained from a simple rotational diagram analysis is superposed in red.
\label{fig:figure4}
}}
\par\end{center}
\label{correl}
\end{figure}

The full coverage of the band 80 -- 350 GHz with ASAI has allowed us to detect towards B1 nine transitions of the type $J_{0,J}$--$J^{\prime}_{0,J^{\prime}}$, with $J^{\prime}= J$ --1, from $J= 4$ up to $J= 15$ ($E_u = 126.6\K$).  Towards B2, we could detect four transitions, between $J= 4$ and $J= 12$. The line parameters and their fluxes are summarized in Table~\ref{tab:table1} and~\ref{tab:table2}. A montage of some of the lines detected towards B1 and B2 is displayed in Fig.~\ref{fig:figure4}. 

The emission of the three transitions of HNCO, $4_{0,4}-3_{0,3}$, $5_{0,5}-4_{0,4}$ and $6_{0,6}-5_{0,5}$ was detected for the first time by Rodr\'iguez-Fern\'andez et~al. (2010) towards  the outflow shocks B1 and B2 and the driving protostar L1157-mm.
The intensity of the transitions $4_{0,4}-3_{0,3}$~and~$5_{0,5}-4_{0,4}$  is consistent with those measured by Rodr\'iguez-Fern\'andez et~al. (2010) towards all three sources. Contrarily to us, the transition $6_{0,6}-5_{0,5}$ was detected by these authors only towards L1157-mm. Despite thorough inspection and re-analysis of their data, no explanation could be found to the lack of detection towards B1 and B2.
We stress that we detected emission from 6 transitions higher than the $6_{0,6}-5_{0,5}$ and as high as $15_{0,15}$--$14_{0,14}$ ($E_u = 126.3\K$) with a very good signal to noise ratio towards B1. Therefore, the physical and chemical properties of HNCO must be revisited, and analyzed based on the broad range of information at hand.

As can be seen in Fig.~\ref{fig:figure2}(b), the profiles of the low-excitation transitions of HNCO can be fit with an intensity-velocity distribution law $I(v) \propto \exp(-|v/v_0|)$, with $v_0=~2.5~\kms$,  similar to the signature of component g3. This suggests a common origin to the low-excitation line emission of NH$_2$CHO and HNCO.

\subsection{Physical conditions}
\begin{figure}
\begin{center}
\includegraphics[width=8.7cm,keepaspectratio]{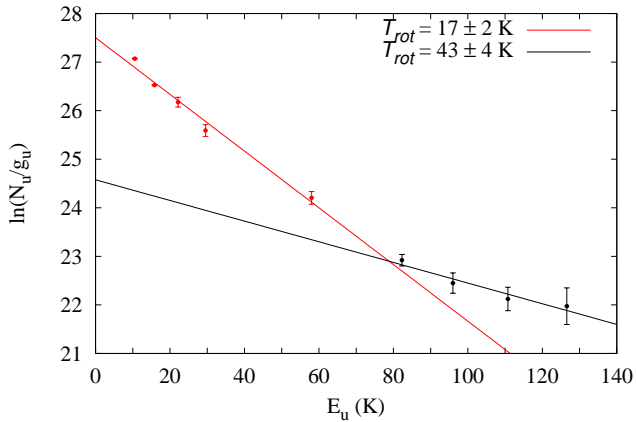}
{\caption{\small Rotational diagram of the HNCO emission detected towards B1. The red and black straight lines display the fits to the components of low- ($T_{rot}= 17\K$) and high- ($T_{rot}= 43\K$) excitation, respectively.
\label{fig:figure5}
}}
\par\end{center}
\end{figure}

In a first step, we have carried out a simple rotational diagram analysis of the HNCO emission, similar to that applied to formamide. The rotational diagram is displayed in Fig.~\ref{fig:figure5}, and again suggests the presence of two components towards B1: a low-excitation component with  $T_{rot}= 17\pm2\ \K$ and $N \rm (HNCO)= 3.3\pm0.5 \times 10^{13}\cmmd$, and a component of higher excitation with $T_{rot}= 43\pm4\ \K$ and column density $N\rm (HNCO)= 8.4\pm1 \times 10^{12}\cmmd$. The very good fit suggests that the lines are close to being thermalized. A similar LTE analysis of the emission towards B2 yields $T_{rot} = 13\pm3\ \K$ and $N\rm (HNCO)= 4.6\pm0.9 \times 10^{13} \cmmd$. The column densities of HNCO derived are higher by a factor of~2~compared to those derived in Rodr\'iguez-Fern\'andez et~al. (2010), reaffirming their conclusion that the HNCO abundances in the B1 and B2 shocks are among the highest HNCO abundances ever measured. We note that the authors failed to detect the warmer, gas component towards~B1.

An important issue to resolve in order to understand the origin of the HNCO emission is whether the two excitation temperatures identified in the rotational diagram arise from non-LTE excitation conditions or actually trace the presence of two physically distinct components. Unlike the case of formamide, approximate collisional excitation rates are available for HNCO in the range 30 to $250\K$. We have analyzed the emission of HNCO using Madex, a radiative transfer code in the Large-Velocity Gradient (LVG) approximation (Cernicharo et al. 2012). In order to minimize the variations of the beam-source coupling across the range of transitions observed, we have estimated the flux of the different transitions in a telescope beam of $20\arcsec$ (HPFW), following the same method as for CO and H$_2$O (Lefloch et~al. 2012; Busquet et~al. 2014). It comes out that the emission of the $4_{0,4}-3_{0,3}$ and $5_{0,5}-4_{0,4}$ transitions can be accounted for only by gas at kinetic temperatures less than about $30\K$.  In fact, we find that the emission of all the transitions up to $7_{0,7}$--$6_{0,6}$ is well accounted for by a single gas component at kinetic temperature $T_{kin}\simeq 22\K$, density $n(\htwo)= 6\times 10^4\cmmt$ and source-averaged column density $N \rm (HNCO)= 3.3 \times\ 10^{13}\cmmd$. A second, warmer gas component ($T_{kin}\approx 65\K$), of denser gas $n(\htwo)\simeq 1.0\times 10^6\cmmt$ and source-averaged column density
$N \rm (HNCO)= 6.5 \times 10^{12}\ \cmmd$ is required to account for the emission of the higher$-J$ transitions. Both components are extended with a typical size of $18-20\arcsec$, similar to that of the CO components g2 and g3. These values are in very good agreement with a simple LTE analysis.

As a conclusion, our LVG analysis of the HNCO transitions confirms that two physical components with different excitation conditions contribute to the emission detected. The physical conditions we derive for both components are similar to those of g3 (the B2 cavity) and g2 (the B1 cavity), as identified from CO (Lefloch et~al. 2012). This interpretation is supported by the similarity of the profiles of the low-$J$ transitions of HNCO with g3.

\subsection{Isomers}

HNCO has several isomers: HCNO, HOCN and HONC. As noticed by Marcelino et~al. (2009), the chemical pathways leading to these different isomers could be very different. Their detection and their relative abundance can provide keys to their formation route. Marcelino et~al. (2009) showed that the relative abundances of HNCO and HCNO in dark clouds could be accounted for by gas phase formation in dark clouds, where $\rm [HNCO] \simeq (1-5)\times 10^{-10}$.

\begin{figure}
\begin{center}
\includegraphics[width=8cm,keepaspectratio]{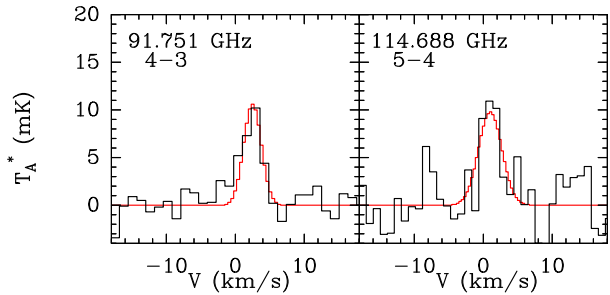}
{\caption{Montage of the HCNO $J$ = 4 -- 3 and $J$ = 5 -- 4 spectra observed towards L1157-B1. The LTE fit to the emission obtained from a simple rotational diagram analysis is superposed 
in red.}
\label{fig:figure6}}
\end{center}
\end{figure}

Searching for HCNO emission towards B1, we detected the transitions $J=$ 4 -- 3 at 91751.32 MHz and $J=$ 5 -- 4 at
114688.38 MHz, respectively. Lines are weak, with typical intensities of $\approx 10$ mK ($T_{\rm A}^{*}$). This is the first detection of HCNO in a protostellar shock region.  The line parameters are given in Table~\ref{tab:table1} and the spectra are shown in Fig.~\ref{fig:figure6}.

From a simple rotational diagram analysis, we derive  a rotational temperature and column density of $10\pm3\K$ and $6.6\pm0.9\times 10^{11}\cmmd$, respectively. This corresponds  to an abundance $\rm [HCNO] \simeq 6.6\pm0.9\times 10^{-10}$.   Under the conservative assumption that the emission of both HCNO and HNCO arise from the same region and are dominated by the low-excitation component g3, we estimate a relative abundance ratio R= [HNCO]/[HCNO] $\thickapprox$ 60. The overall abundances of NH$_2$CHO, HNCO, H$_2$CO and CH$_3$OH are summarized in Table~\ref{tab:table3}.

\section{Acetamide}
Acetamide, NH$_2$COCH$_3$, is the next most simple amide after formamide. Acetamide has been detected only towards SgrB2(N) until now (Hollis et~al. 2006; Halfen, Ilyushin \& Ziurys 2011). In that source,  Halfen, Ilyushin \& Ziurys~(2011) report acetamide abundances comparable to, or even higher than that of formamide, of the order of $(0.2-2.0)\times 10^{-10}$.  Because of the importance of acetamide and formamide in prebiotic chemistry, we also searched for acetamide towards B1 and B2. In spite of the excellent sensitivity of our survey, our search  has yielded only negative results.

This species displays an extremely rich spectrum at millimeter wavelengths. The observations by Halfen, Ilyushin \& Ziurys~(2011) in SgrB2(N) show that, in the millimeter range, a few transitions appear to be free of contamination by other molecular transitions. In the ASAI survey of B1, we have selected  the $8_{0,8}-7_{0,7}$ transitions 87632.435 MHz among such transitions, as it arises from a level with a rather low $E_u$~($19.80\K$), while the Einstein spontaneous emission coefficient is rather high. We have assumed an excitation temperature $T_{exc}\simeq 8\K$, similar to the values derived for NH$_2$CHO and HNCO in the low-excitation gas which dominates the emission (Table~\ref{tab:table3}).

At the frequency of NH$_2$COCH$_3$ $8_{0,8}-7_{0,7}$, we measure an rms noise of  $\approx 1.5$~mK in a velocity interval of $1.3\kms$. Under these assumptions, we obtain a $3\sigma$ upper limit of $4\times 10^{-10}$ for $\rm [NH_2COCH_3]$. Hence, the  upper limit on the ratio $\rm [NH_2COCH_3]/[NH_2CHO]$ is $\approx 0.15$. Our results indicate that acetamide is much less abundant than formamide, in contrast to what was found in SgrB2(N).

\begin{table}
\caption{Abundances of NH$_2$CHO, HNCO, H$_2$CO and CH$_3$OH in B1 and B2, with respect to H$_2$. Methanol and  Formaldehyde abundances are taken from Lefloch \& Bachiller (in preparation). 
\label{tab:table3}}
\begin{tabular}{lcccc}
\hline
Source   & $\rm [NH_2CHO]$    & $\rm [HNCO]$       & $\rm [H_2CO]$  &  $\rm [CH_3OH]$ \\
     &10$^{-8}$&10$^{-8}$&10$^{-7}$    &10$^{-6}$       \\    \hline
B1: g3 & $0.35\pm 0.06$  & $3.3\pm 0.5$   &    $\cdots$ & $\cdots$ \\
B1: g2 & $0.17\pm 0.04$  & $0.84\pm 0.1$  &    $\cdots$ & $\cdots$ \\
B1     & $0.52\pm 0.1$  & $4.1\pm 0.6$    &   $3.0\pm 1$  &  $1.5\pm 5$\\
B2     & $1.1\pm 0.2$  & $8.5\pm 2$   &   $4.0\pm 2$  & $2.5\pm  6$\\
\hline
\end{tabular}
\end{table}

\section{Origin of Formamide and Isocyanic Acid}

\subsection{The correlation between HNCO and NH$_2$CHO}

\begin{figure}
\begin{center}
\includegraphics[width=0.8\columnwidth,keepaspectratio]{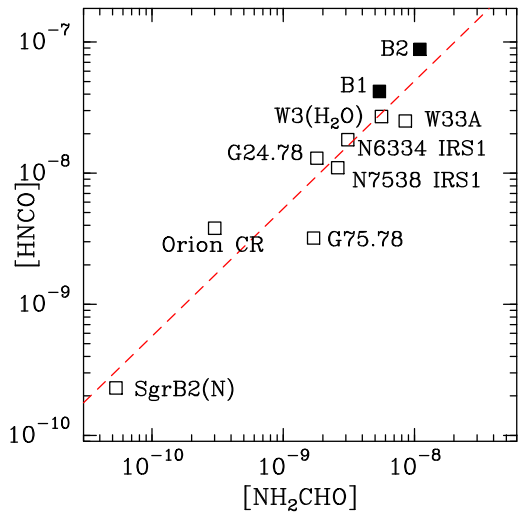}
{\caption{Comparison of molecular abundances for NH$_2$CHO and HNCO in a sample of Galactic sources. Filled symbols indicate our values measured towards the protostellar shocks B1 and B2. The  data were taken from by Sutton et~al. (1995), Bisschop et~al. (2007) and Halfen, Ilyushin \& Ziurys~(2011). The best power-law fit $\rm [HNCO]= 3.0\times [NH_2CHO]^{0.97}$ is drawn in dashed.
\label{fig:figure7}
}}
\par\end{center}
\label{correl}
\end{figure}

We report in Fig.~\ref{fig:figure7} the abundance of NH$_2$CHO together with that of HNCO in B1 and B2, and in the various Galactic regions where these species were detected (Sutton et al. 1995; Bisschop et~al. 2007; Halfen, Ilyushin \& Ziurys 2011).

We find that NH$_2$CHO and HNCO are as abundant, or even more abundant towards protostellar shocks B1 and B2 than in the richest Galactic sources known until now, like e.g. W3A or W3(H$_2$O). A tight linear correlation is observed between the NH$_2$CHO and HNCO abundances, with a Pearson correlation coefficient $r=0.93$. A power-law fit yields $\rm [HNCO]\simeq 2.84\times [NH_2CHO]^{0.97}$ (see Fig.~\ref{fig:figure7}).
Hence,  NH$_2$CHO and HNCO are tightly chemically related species.  More observations are needed to confirm this correlation on a firmer statistical basis. This suggests that either they form from a common precursor (\lq\lq parent\rq\rq) molecule, or that one forms from the other. Therefore, studying the formation route of HNCO can shed light on the formation pathway of formamide itself.

\subsection{Isocyanic acid}
The formation of HNCO has attracted a lot of attention both in dark clouds and hot cores. Some authors have considered formation only in
the gas phase (e.g. Iglesias 1977; Turner, Terzieva \& Herbst 1999) while others have modelled the formation on grain surfaces
(D'Hendecourt \& Allamandola 1986; Hasegawa \& Herbst 1993; Garrod, Weaver \& Herbst 2008; Tideswell et~al. 2010).  The possible formation pathways of HNCO in shocks have never been modeled.
Rodr\'iguez-Fern\'andez et~al. (2010) have investigated in detail the efficiency of the gas phase formation routes proposed by Tideswell et~al. (2010) when applied to shocks physical conditions.
These authors conclude that the very high HNCO abundances measured in the young shocks B1 and B2 can not be accounted for by pure gas phase reactions alone, and that formation on dust grain mantle followed by sputtering in the shock is the most likely explanation.

\subsection{Formamide}
Several reaction pathways have been previously proposed to explain the abundance of gas-phase formamide in the interstellar medium: formamide might form either on the icy mantles of dust grains or directly in the gas phase. The actual reaction pathways and their efficiency remain uncertain.

\subsubsection{Gas phase formation}

In their study of the solar-type protostar IRAS16293-2422, following the KIDA database\footnote{http://kida.obs.u-bordeaux1.fr/} (Wakelam et~al. 2012), Kahane et~al. (2013) investigated the formation of formamide in the gas phase from the exothermic reaction between formaldehyde H$_2$CO and amidogen NH$_2$:
\begin{equation*}
{\rm NH_2 + H_2CO \rightarrow NH_2CHO + H\ \ \ (I)}
\end{equation*}
while it is removed from the gas phase mainly through reaction with HCO$^{+}$
\begin{equation*}
{\rm NH_2CHO + HCO^{+} \rightarrow CO + NH_2CH_2O^{+}\ \ \ (II)}
\end{equation*}
Adopting a value of $10^{-10}$ at $100\K$ for the rate of reaction~(I), typical of barrierless, neutral-neutral reactions, the authors could successfully account for the gas phase abundance of formamide towards the solar-type protostar IRAS16293.

Based on these results, we have investigated whether the formation pathway proposed by Kahane et~al. (2013) could also account for the formamide abundances measured in the protostellar shocks B1 and B2.
To do so, we have implemented reactions (I) and (II) with the rates proposed by Kahane et~al. (2013) in the time-dependent gas phase-chemistry code Astrochem\footnote{http://smaret.github.io/astrochem/},
in order to (i)~compute the chemical composition of the parental cloud, (ii)~determine whether the change of physical conditions induced by the shock crossing are sufficient to produce a high enough abundance of formamide. The code reads a network of chemical reactions, builds a system of kinetic rates equations, and solves it using a state-of-the-art stiff ordinary differential equations solver. It considers gas-phase
processes and simple gas-grain interaction, such as freeze-out, and desorption via several mechanisms (thermal, cosmic-ray,
and photodesorption). In our calculations we use the OSU chemical network\footnote{http://www.physics.ohio-state.edu/~eric/research.html}, we assume a visual extinction $A_V = 10$ mag consistent with \ceio\ and \thco\ column densities, a grain size of $0.1~\mu m$, and initial elemental abundances as assumed by Wakelam, Herbst \& Selsis (2006) in their low-metallicity model with the exception of He and N for which we assume cosmic abundances (e.g. Asplund et~al. 2005; Tsamis \& Walsh 2011).

\begin{figure}
\begin{center}
\includegraphics[width=0.9\columnwidth,keepaspectratio]{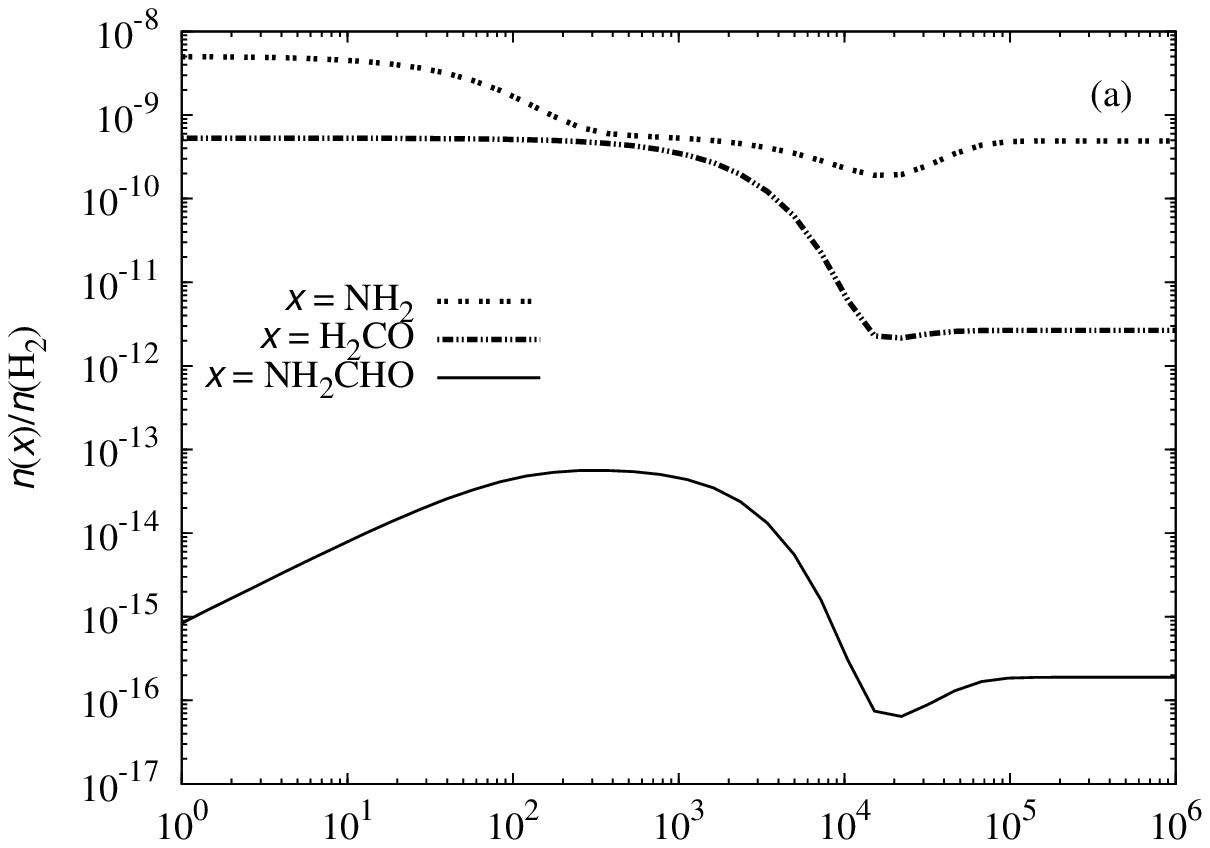}
\includegraphics[width=0.9\columnwidth,keepaspectratio]{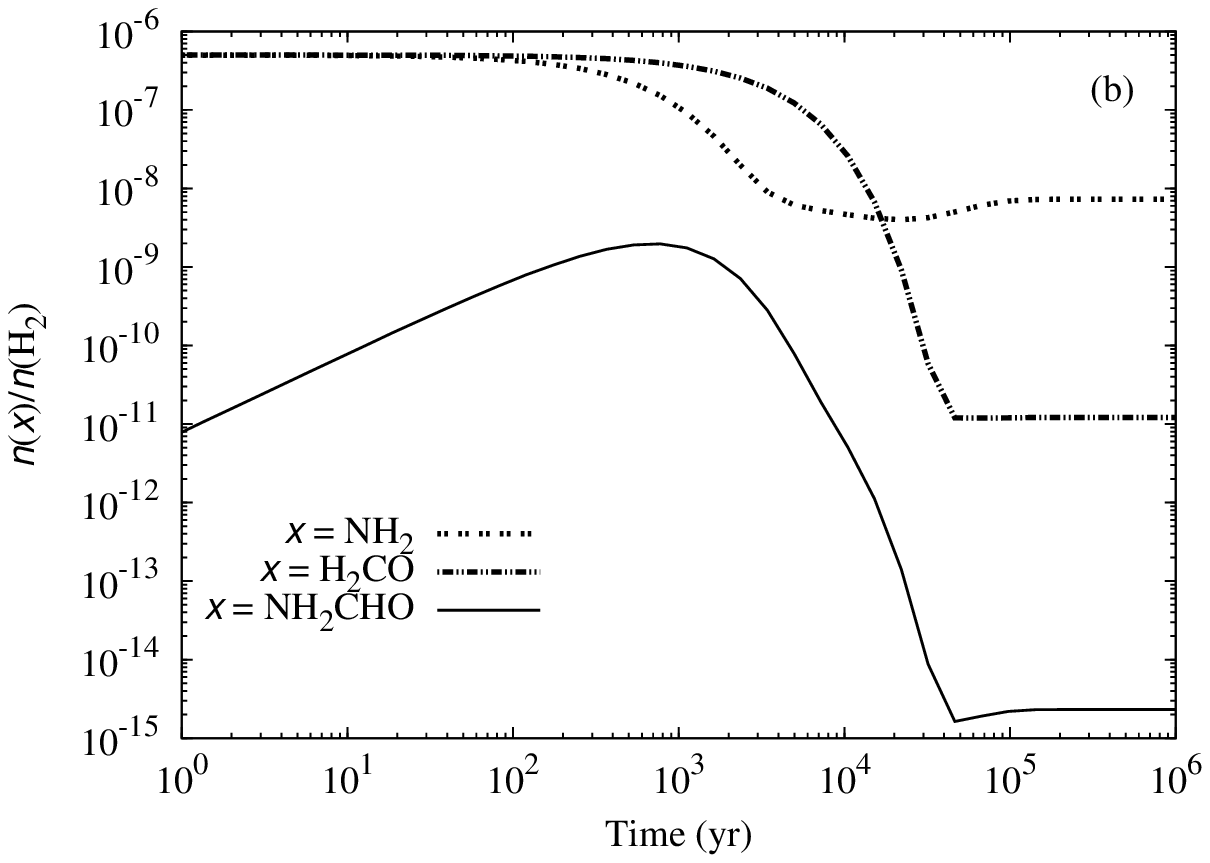}
{\caption{Modelling of formation of formamide from formaldehyde and amidogen (reaction I in the text). Time variation of $\rm [NH_2CHO]$ (solid line), formaldehyde $\rm [H_2CO]$ (dotted-dashed line) and amidogen $\rm [NH_2]$ (dashed line), after passage of the shock.
{(a)}~without grain sputtering: $T=70\K$, $n(\htwo)= 10^5\cmmt$; {(b)}~with grain sputtering: $T=70\K$, $n(\htwo)= 10^5\cmmt$, $\rm [H_2CO]= [NH_2]= 5\times 10^{-7}$.
\label{fig:figure8}
}}
\par\end{center}
\end{figure}

In a first step, we have computed the gas chemical composition in the cloud, at steady-state, before the passage of the shock. Not surprisingly, we found very low abundances $\rm [NH_2CHO]$, of the order of a few  $10^{-15}$, in the cloud gas, with a kinetic temperature $T_{kin}=10\K$ and density $n(\htwo)= 10^4\cmmt$. The predicted abundances of formaldehyde and NH$_2$ are $5\times 10^{-10}$ and $5\times 10^{-9}$, respectively.

In a second step, we have computed how the steady-state molecular abundances evolve after the passage of the shock, when the gas is compressed in the outflow cavity walls to a density of about $10^5\cmmt$ and the temperature rises up to $70\K$ (see e.g. Lefloch et~al. 2012; Podio et~al. 2014). Here, we did not take into account any sputtering of H$_2$CO and NH$_2$ from the grain mantles in the passage of the shock. As can be seen in Fig.~\ref{fig:figure8}(a), it comes out that both the abundances of  NH$_2$CHO and H$_2$CO (10$^{-14}$ and $7\times 10^{-9}$, respectively) are still too low after the passage of the shock. Therefore, a pure gas phase scenario does not work.

We have then considered the possibility that formaldehyde and amidogen  actually formed on the mantles of dust grains and were subsequently released in the gas phase through sputtering of the icy grain mantles. The detection of bright methanol and SiO emission lines indicates that sputtering is indeed very efficient in the shock regions B1 and B2. We have modelled the sputtering process by rising suddenly the abundance of formaldehyde and amidogen together with the density and kinetic temperature of the cloud. In particular, [H$_2$CO] was risen up to its observed value ($5\times 10^{-7}$). As can be seen in Fig.~\ref{fig:figure8}(b), it is necessary to  rise the abundance of amidogen up to $\rm [NH_2]= 5\times 10^{-7}$ at the passage of the shock so that [NH$_2$CHO] increases up to $\approx 10^{-9}$ at $\approx 10^3\yr$.  This is still a factor of 10 less than the abundance measured towards B2.

Furthermore, the ground transition of NH$_2$  at 952.578354~GHz was not detected by Herschel in the CHESS key-program (Ceccarelli et~al. 2010). The observations allowed us to reach a rms noise of $\approx 6$~mK per velocity interval of $1\kms$ (Obs\_ID: 1342201555 and 1342207324). Adopting an excitation temperature $T_{exc}=10\K$, this allows to place an upper limit of a few $10^{-10}$ on the abundance of amidogen; this is at least two orders of magnitude less than the abundance predicted by our gas phase model at the formamide abundance peak.

Observations with Spitzer and Herschel reveal the presence of temperatures as high as $\sim 10^3\K$ in the shock (Nisini et al. 2010; Busquet et al. 2014), as expected for magnetized shocks (Flower \& Pineau Des For\^ets 2010).  Using Astrochem, we have investigated the impact of such a high-temperature region on the formation of formamide, by adopting a gas temperature in the range $200\K$ -- $1000\K$  in the second step of the simulation. We found that the variation of formamide abundance with time is essentially unchanged with respect to the previous results, when the post-shock temperature remains $\simeq 500\K$. At higher temperatures ($1000\K$), the production of formamide is less efficient: the formamide abundance  rises a maximum of a few $10^{-11}$ after $\approx 100\yr$ and subsequently decreases. Therefore, the high temperatures reached by the gas in the passage of the shock does not enhance the production of formamide.

\begin{figure}
\begin{center}
\includegraphics[width=0.8\columnwidth,keepaspectratio]{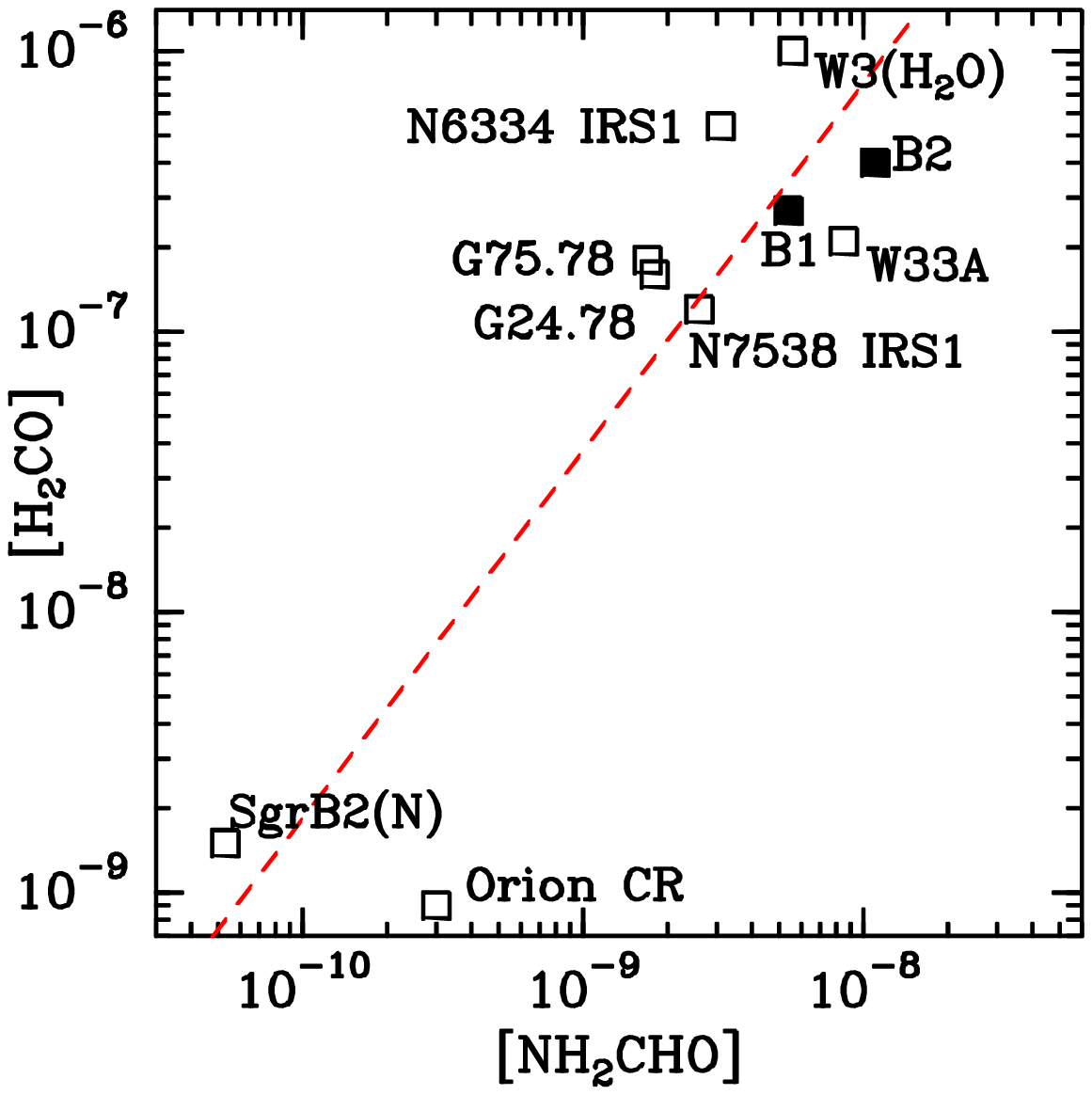}
{\caption{Comparison of molecular abundances for NH$_2$CHO and H$_2$CO. Filled symbols indicate our values measured towards the protostellar shocks B1 and B2. The  data were taken from by Sutton et~al. (1995), Bisschop et~al. (2007) and Halfen, Ilyushin \& Ziurys~(2011). The best power-law fit $\rm [H_2CO]= 7.8\times10^3 [NH_2CHO]^{1.26}$ is drawn in dashed.
\label{fig:figure9}
}}
\par\end{center}
\label{correl}
\end{figure}

We conclude that even when taking account the sputtering of dust grain mantles in shocks, the formation pathway from formaldehyde and amidogen fails to account for the abundance of formamide in B1 and B2. These results are consistent with the poor linear correlation observed in Fig.~\ref{fig:figure9} (r= 0.70) between the abundances of NH$_2$CHO and H$_2$CO in Galactic sources.

Theoretically, Li \& L\"u (2002) and Redondo, Barrientos \& Largo~(2014) have noticed that the first step of the reaction pathway between  H$_2$CO and NH$_2$ actually has an activation barrier of a few $10^3\K$, which significantly lowers the efficiency of the reaction even in shock regions. Li \& L\"u (2002) also have shown that the production of HCO and NH$_3$ is actually favored over NH$_2$CHO and H.
More generally, as for the gas phase pathways proposed for formamide, Jones, Bennett \& Kaiser~(2011) noticed that most of the reactions involved between neutral species have activation barriers significantly higher than the kinetic energy available even in the hot core regions of star-forming molecular clouds. Redondo, Barrientos \& Largo~(2014) have investigated different gas-phase reactions between cations with a nitrogen atom (e.g. NH$_4^{+}$) and neutral molecules with a carbonyl group (H$_2$CO and HCOOH), that could produce precursors of formamide. The analysis of the potential energy surfaces corresponding to these reactions shows that these processes present a net activation barrier.
Hence, there is presently no gas phase reactions able to account for the abundances of formamide measured in B1 and B2.

\subsubsection{Grain chemistry}

Formamide has been tentatively assigned within the solid phase on icy grains toward the protostellar objects NGC 7538 IRS9 and W33A.

Quantifying the amount of formamide present in the cold envelope of these young stellar objects is difficult due
to the broad overlapping features with, for instance, ammonia. The amount of solid formamide believed to be present within these young protostellar objects is therefore unknown.

Theoretical and experimental works have proposed several reaction pathways for the formation of formamide on grain mantles. For instance,
Garrod, Weaver \& Herbst~(2008) investigated the formation of complex organic molecules on the ice mantles of dust grains, from reactions involving radicals. These  radicals are produced primarily by cosmic ray-induced photodissociation of ices formed during the cold, early stages of evolution. In this model, formamide can form in ices by hydrogenation (addition of H atoms) starting with OCN, and also by the reaction of HCO and amidogen. This model is very efficient at producing formamide, which is predicted overabundant ($\rm [NH_2CHO] \approx$ a few $10^{-6}$), whereas isocyanic acid is predicted quite underabundant ([HNCO]~$\approx 10^{-9}$).

Laboratory experiments on electron and vacuum UV (VUV) irradiation of a mixture of CO-NH$_3$ ices have identified formamide as a reaction product (Demyk et~al. 1998; also Jones, Bennett \& Kaiser 2011 for a review).

Raunier et~al. (2004) identified formamide as one of the products of the UV irradiation of pure ices of HNCO at 10~K.
They proposed that the formation of formamide proceeds from hydrogenation of HNCO, as follows:
\begin{eqnarray*}
\rm HNCO + H \rightarrow NH_2CO\ \ \ (III) \\
\rm NH_2CO + H \rightarrow NH_2CHO\ \ \ (IV){.}
\end{eqnarray*}

Our observations bring some interesting constraints on the formation route of formamide, when comparing the amount of material released from dust  grains in B1 and in B2. Methanol (CH$_3$OH) can be used as a probe of the amount of material formed on the dust grain mantles and released in the gas phase from sputtering in the shock. It is widely accepted that methanol is synthesized on grain surfaces (Tielens \& Hagen 1982) as gas phase reactions are unable to reproduce the large observed abundances (Roberts, Herbst \& Millar 2004; Maret et~al. 2005). This is supported by recent experimental and theoretical works, which have confirmed the synthesis of methanol by hydrogenation of iced CO (Watanabe \& Kouchi 2002; Hidaka et~al. 2008; Rimola et~al. in preparation).  Methanol line emission in B1 and B2 has been recently analyzed by Lefloch et~al. (in preparation), who determined abundances of $1.5\pm0.5\times 10^{-6}$ and $2.5\pm0.6\times 10^{-6}$, respectively.

Hence, as can be seen in Table~\ref{tab:table3}, the abundance variations of HNCO, NH$_2$CHO and CH$_3$OH are similar and remain modest. The lack of marked variations in the methanol abundance  is in agreement with the idea that the dust grain mantle properties and sputtering efficiency being similar at both positions. The abundance enhancements of isocyanic acid and formamide between B1 and B2 are in quantitative agreement with the variations of the amount of material present on the grains, as traced by methanol. Hence, we conclude that HNCO and NH$_2$CHO most likely formed on dust grains and were released in the gas phase from sputtering in the shock.

The linear correlation observed between the abundances of isocyanic acid and formamide favors the scenario proposed by Raunier et~al. (2004), among the various formation pathways proposed for formamide, in which NH$_2$CHO forms from subsequent hydrogenations of HNCO on mantle ices. However, unlike the experiment of these authors, there is no source of UV irradiation in the shock regions B1 and B2.

More theoretical work is needed to investigate the efficiency of the hydrogenation process of HNCO on dust grains, in the conditions typical of cold molecular clouds. Also, it remains  to understand the formation pathway of the HNCO itself, whether it formed from hydrogenation of OCN, as proposed by Garrod, Weaver \& Herbst~(2008), or through a different process.

\section{Conclusion}
We have carried out the first systematic study of molecules with a peptide link in protostellar shocks, as part of the ASAI Large Program.
\begin{itemize}
\item We have detected formamide, towards the shock regions B1 and B2 in the outflow of the Class 0 protostar L1157-mm. Our spectral surveys have allowed us to detect  transitions with a wide range of $E_u =$  (10-- $80\K$) towards B1 and B2.  A search for the next most simple amide, acetamide NH$_2$COCH$_3$, yielded negative results. Emission from HNCO  was also detected in transitions as high as $15_{0,15}$ -- $14_{0,14}$ at 329.664 GHz, with $E_u= 126.6\K$.  The isomer HCNO could be detected towards B1.

\item The excitation conditions of formamide and isocyanic acid were derived in the LTE approximation. Towards B1, we find that two components of low- and high-excitation, respectively, are contributing to the emission. A modelling of the HNCO emission towards B1 with a radiative transfer code in the Large Velocity Gradient approximation shows that the physical conditions of both components are very similar to those derived from CO for the B1 outflow cavity (g2) and the B2 outflow cavity (g3)  (Lefloch et~al. 2012; G\'omez-Ruiz et~al. submitted). These results are supported by the similarity between the HNCO and NH$_2$CHO line profiles and those of CO.

\item Towards B1, we estimate for g3 gas column densities  $N(\rm NH_2CHO)= 3.5\pm0.6\times 10^{12}\cmmd$ and $N(\rm HNCO)= 3.3\pm0.5\times 10^{13}\cmmd$, and molecular abundances $\rm [NH_2CHO]= 3.5\pm0.6\times 10^{-9}$ and $\rm [HNCO]= 3.3\pm0.5\times 10^{-8}$. For the g2 component, we obtain $N(\rm NH_2CHO)= 1.7\pm0.4\times 10^{12}\cmmd$ and $N(\rm HNCO)= 8.4\pm1\times 10^{12}\cmmd$ corresponding to molecular abundance $\rm [NH_2CHO]= 1.7\pm0.4\times 10^{-9}$ and $\rm [HNCO]= 8.4\pm1\times 10^{-9}$.  Towards B2, we estimate a gas column density $N(\rm NH_2CHO)= 5.9\pm1.4\times 10^{12}\cmmd$ of low excitation $ T_{rot}=10\pm3\K$, and $N(\rm HNCO)= 4.6\pm0.9\times 10^{13}\cmmd$; and abundances about $\rm [NH_2CHO]= 1.1\pm0.2\times 10^{-8}$ and $\rm [HNCO]= 8.5\pm2\times 10^{-8}$.

\item Comparison with previous studies on isocyanic acid and formamide shows that both molecules are more abundant towards protostellar shocks B1 and B2 than in the richest Galactic sources known so far. A tight linear correlation between HNCO and NH$_2$CHO is observed, which suggests that both species form from the same precursor or that one form from the other. The variations of HNCO and NH$_2$CHO abundance between B1 and the older shock B2, are moderate and found consistent with those of methanol, a tracer of the material in the icy mantles of dust grains, which supports the idea that these molecules were probably released in the gas phase from dust mantle sputtering in passage of the shocks.

\item The abundance of formamide and HNCO in the gas phase cannot be accounted for by the present gas-phase models, in particular from the reaction between formaldehyde and amidogen. From comparison with methanol, we favor the scenario proposed by Raunier et~al. (2004), in which NH$_2$CHO forms from subsequent hydrogenation of HNCO on dust grains. 
\end{itemize}

\section*{Acknowledgments}
Based on analysis carried out with the CASSIS software and the JPL and CDMS spectroscopic databases. CASSIS has been developed by IRAP-UPS/CNRS.
E. Mendoza thanks support from COSPAR during his stay at IPAG. We thank the French PCMI for financial support. We thank  the INEspa\c{c}o, CNPq, CAPES and FAPERJ Brazilian institutions for the advancement research. Rafael Bachiller gratefully acknowledges partial support from
Spanish MINECO Grant FIS2012-32096. We thank W. Bethune, N. Guillard, M. Keppler, A. Wunsch for help with the observations.

\label{lastpage}

\end{document}